# Field-resolved observation and quantification of the instantaneous Kerr response in semiconductors

Haifeng Lin[1¶*], Hannes Kempf[1¶*], Hinduja Bhuvanendran[1], Aitor De Andrés[1], Markus Ludwig[2,3], and Nicolò Maccaferri[1,4,5*]

[1]Ultrafast Nanoscience Group, Department of Physics, Umeå University, 901 87 Umeå, Sweden
[2]Department of Physics and Materials Science, University of Luxembourg, L-1511 Luxembourg, Luxembourg
[3]Institute for Advanced Studies, University of Luxembourg, L-4365 Esch-sur-Alzette, Luxembourg
[4]Umeå Centre for Microbial Research, Umeå University, 901 87 Umeå, Sweden
[5]Wallenberg Initiative Materials Science for Sustainability, Department of Physics, Umeå University, 901 87 Umeå, Sweden
[*]Address correspondence to: lhf.yurnhero@gmail.com, hannes.kempf@umu.se, nicolo.maccaferri@umu.se
[¶]Contributed equally

**Abstract**
The optical Kerr effect, whereby the refractive index of a material changes with light intensity, underpins nonlinear photonics and emerging mid-infrared technologies. Accurate determination of the nonlinear refractive index is therefore essential, yet conventional intensity-based techniques infer the nonlinear phase shift indirectly from changes in beam propagation or optical spectra. Moreover, methods such as the z-scan cannot resolve the temporal evolution of the nonlinear response, making it difficult to distinguish instantaneous Kerr nonlinearities from delayed contributions, including free-carrier effects. Here, we investigate the nonlinear optical response of germanium at 5 μm using field-resolved degenerate pump-probe spectroscopy. By measuring the transmitted probe electric field as a function of pump-probe delay, we directly retrieve the pump-induced phase shift and its temporal evolution while retaining sensitivity to changes in absorption. We observe cross-phase modulation that follows the pump intensity envelope and vanishes when the pump and probe no longer overlap, consistent with an instantaneous Kerr response. The phase shift scales linearly with pump intensity, yielding a nonlinear refractive index of $n_2 = (0.87 \pm 0.16) \times 10^{-17}\,\mathrm{m^2/W}$. Our results establish the instantaneous Kerr effect as the dominant contribution to the measured nonlinear phase shift under our experimental conditions and demonstrate field-resolved pump-probe spectroscopy as a direct, time-resolved method for determining the nonlinear refractive index and distinguishing instantaneous from delayed optical responses.

## INTRODUCTION

The optical Kerr effect, arising from the third-order nonlinear response of a medium, produces an intensity-dependent refractive index. For a propagating optical field, this intensity dependence gives rise to a nonlinear phase shift, whose magnitude is governed by the nonlinear refractive index $n_2$. This makes $n_2$ a fundamental parameter for nonlinear optical design, determining the strength of self- and cross-phase modulation and governing phenomena such as self-focusing and soliton formation [1]. In the mid-infrared, the Kerr effect underlies a range of nonlinear optical applications, including Kerr-lens mode locking of oscillators [2, 3], nonlinear pulse compression [4, 5], octave-spanning supercontinuum generation [6, 7], and Kerr soliton formation in microresonators [8, 9]. Quantitative knowledge of $n_2$ at mid-infrared wavelengths is therefore essential for predicting and controlling the nonlinear phase and the resulting evolution of intense optical fields.

A variety of techniques have been developed to measure nonlinear refraction and determine $n_2$. The Z-scan infers nonlinear refraction from the intensity-dependent evolution of a focused beam [10–13], degenerate four-wave mixing measures the energy diffracted by the induced nonlinear polarization [14, 15], time-resolved interferometry measures the nonlinear phase through the displacement of interference fringes [16], spectrally resolved two beam coupling reads the pump-induced reshaping of the probe spectrum [17, 18], and spectral interferometric approaches retrieve the phase response from the frequency-dependent interference of a probe with a reference pulse [19, 20]. These techniques provide complementary access to nonlinear refraction, each probing a different manifestation of the response and requiring its own experimental and analytical procedure to recover $n_2$. Their implementation becomes particularly challenging in the mid-infrared, where suitable detectors are less sensitive and less widely available and broadband femtosecond pulses are more demanding to generate and characterize. Consequently, quantitative measurements of $n_2$ in the mid-infrared remain comparatively sparse, and reported values vary substantially between measurements, even for well-studied optical materials [21, 22].

The central challenge in quantifying the Kerr effect is that the nonlinear response is fundamentally encoded in the phase of the optical field, whereas conventional direct detection measures intensity. In addition, to follow its temporal evolution also requires a pump-probe geometry, in which the pump provides a controlled perturbation and the probe reads out the induced response as a function of delay. In conventional intensity-resolved pump-probe detection (Fig. 1(A), top), the observable is the transmitted probe intensity. A purely refractive response does not change the total transmitted probe energy, but a time-dependent phase can redistribute that energy across the probe spectrum, producing wavelength-dependent intensity changes. Such pump-induced spectral features may also contain coherent contributions in transient-absorption measurements, complicating their interpretation [23, 24]. What distinguishes the available implementations is how the probe is read out. Several of the techniques listed above retrieve the nonlinear phase from intensity-based or interferometric observables [16–20]. By contrast, field-resolved detection takes a different route by recording the probe electric field itself as the primary observable (Fig. 1(A), bottom). The measured waveform retains both the amplitude and the phase of the optical field [25], so the pump-induced response is represented directly as a complex field change rather than inferred from an intensity or a spectral redistribution.

In the work this waveform is recorded by electro-optic sampling (EOS), which is well established across the terahertz and mid-infrared spectral regions [26–28], and has more recently been extended toward the near-infrared and visible [29–31]. By encoding the mid-infrared electric field in the polarization of an optical gate pulse, EOS enables field-resolved detection using conventional photodetectors, circumventing the limited sensitivity and availability of mid-infrared detectors. Field-resolved detection has since been used to study a broad range of ultrafast processes in condensed matter and molecules [32–34]. Representative examples include the build-up of many-body interactions in a photoexcited electron-hole plasma [35, 36], the ultrafast nonlinear response of quasiparticles [37, 38], free-carrier dynamics in photoexcited semiconductors [39, 40], and the coherent vibrational fingerprints of molecules [41]. In all of these examples, the field probes a response that outlasts the exciting pulse: photoinjected free carriers, excited quasiparticles, or vibrating molecules that persist after the pump pulse has gone. For nonlinear refractive measurements, however, the challenge is to isolate the instantaneous Kerr contribution from other pump-induced responses. In previous field-resolved studies of strongly driven solids, the Kerr contribution appeared together with photoinjected-carrier responses and excitation-induced changes in absorption, making it difficult to quantify independently [40]. Complementary attosecond transient-reflection measurements in germanium have revealed the interplay between carrier photoinjection and field-driven electron dynamics in shaping its transient optical response [42]. The instantaneous bound-electron Kerr nonlinearity has not been isolated and quantified on its own by field-resolved detection, in a regime well below the carrier-injection threshold, where carrier-mediated contributions are strongly suppressed.

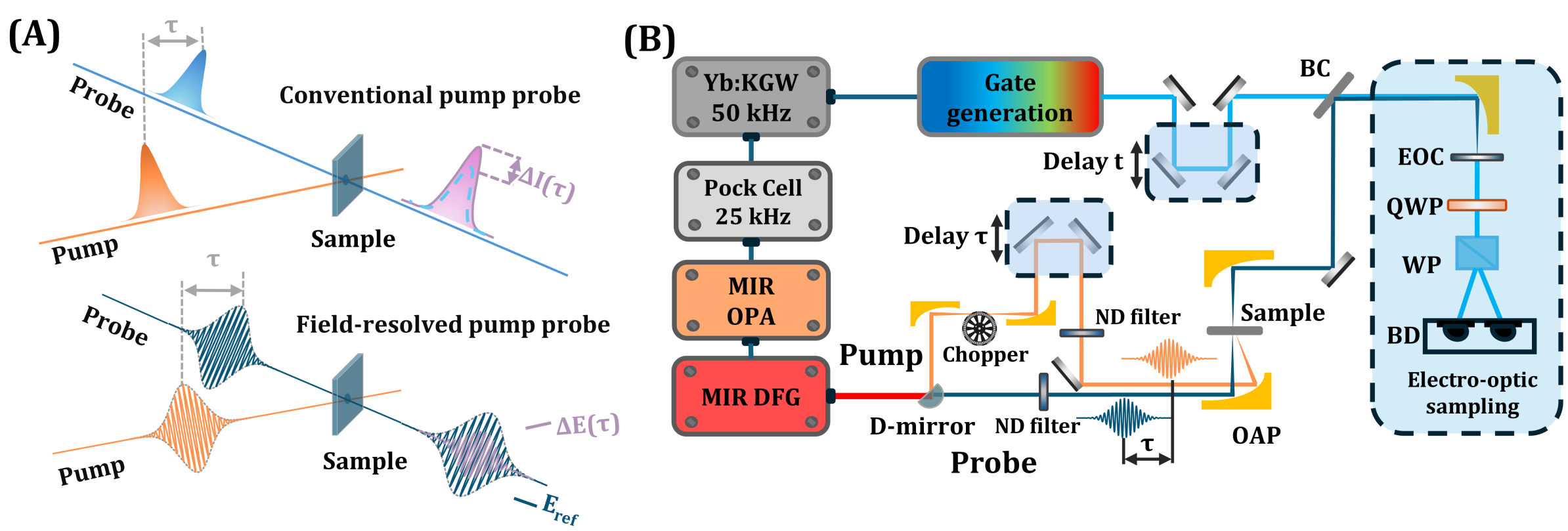

**Fig. 1.** Field-resolved detection of the Kerr response and experimental setup. (A) Comparison of the detected observable in conventional intensity-based pump-probe (top, transmitted-intensity change $\Delta I(\tau)$) and in field-resolved detection (bottom, time-resolved probe field $E_{ref}$ and the pump-induced field change $\Delta E(\tau)$, which carries both the amplitude and the phase of the response). (B) Experimental setup for field-resolved cross-phase-modulation of germanium in the mid-infrared. A single mid-infrared pulse train is split into an identical pump and probe. The pump intensity is varied with neutral density (ND) filters, and a further ND filter attenuates the probe to a weak-probe level. The pump-probe delay τ and the EOS time t are set by two independent delay stages. After the sample, the transmitted probe is combined with the gate pulse using a beam combiner (BC) and sampled in the electro-optic crystal (EOC). OPA, optical parametric amplification; DFG, difference-frequency generation; OAP, off-axis parabolic mirror; QWP, quarter-wave plate; WP, Wollaston prism; BD, balanced photo detector.

Germanium is a representative case of both the interest and the difficulty: its narrow bandgap of approximately 0.66 eV at room temperature and strong optical nonlinearity make it an important material for mid-infrared nonlinear photonics, with nonlinear refractive indices in the order of $10^{-17}$ m²/W reported or expected at these wavelengths [43, 44]. At 5 μm, the photon energy is approximately 0.25 eV, so multiphoton excitation can generate free carriers at sufficiently high intensity, introducing absorptive and delayed contributions that can obscure the underlying Kerr response [45–47]. Here we measure the nonlinear response of germanium at 5 μm by field-resolved degenerate pump-probe spectroscopy. The pump and probe are derived from the same mid-infrared pulse, with a measured spectral full width at half maximum (FWHM) of 5 THz and a temporal intensity FWHM of approximately 127 fs. The maximum average power of the pump pulses is 0.43 mW at a repetition rate of 25 kHz, corresponding to a pulse energy of approximately 17.2 nJ. The transmitted probe waveform is recorded by EOS at each pump-probe delay, giving directly the pump-induced field change $\Delta E$. Fourier transformation of the measured waveforms yields the transmission change $\Delta T/T$, providing simultaneous access to both the amplitude and phase response. Because pump and probe share the same spectrum they propagate at the same group velocity, so no temporal walk-off arises between them, and the Kerr coefficient obtained is the degenerate one, without conversion from a nondegenerate measurement.

Under our low-intensity excitation conditions, we observe a predominantly refractive response whose delay dependence is confined to the temporal overlap of the pump and probe pulses, with no measurable component persisting after their temporal separation. From the linear scaling of the induced phase with intensity we determine the nonlinear refractive index $n_2 = (0.87 \pm 0.16) \times 10^{-17} \mathrm{m^2/W}$. These measurements establish field-resolved degenerate pump-probe spectroscopy as a direct route to the Kerr index in the mid-infrared.

## MATERIALS AND METHODS

The field-resolved degenerate pump-probe setup is shown in Fig. 1(B). The system is driven by an ytterbium:KGW laser (Light Conversion Pharos) delivering 1030 nm pulses at a repetition rate of 50 kHz. The fundamental is divided into two portions. The larger fraction passes through a Pockels cell that transmits every second pulse to an optical parametric amplifier (Light Conversion Orpheus MIR) followed by difference-frequency generation (Light Conversion Lyra), producing carrier-envelope phase-stable mid-infrared pulses at 5 µm and a repetition rate of 25 kHz. The mid-infrared beam is divided by a silver D-shaped mirror into pump and probe arms.

The probe is strongly attenuated with an optical-density-3 (OD3) neutral density filter to minimize probe-induced nonlinear effects. The pump intensity is varied with a set of neutral-density filters in the pump arm, covering an in-sample peak intensity range of 0.49–2.39 GW/cm² after accounting for the Fresnel loss at the front surface. At the lowest pump intensity (OD2) of $0.49 \pm 0.09$ GW $\mathrm{cm^{-2}}$, the peak nonlinear phase ($\Delta\varphi_0$, extracted as described for Fig. 3(A)) is only 0.075 rad. Because the probe intensity is approximately one order of magnitude lower and its self-induced phase is half the cross-induced one per unit intensity, the probe-induced contribution is estimated to be below 0.0038 rad (<5% of the OD2 signal), supporting the assumption that probe-induced nonlinear effects are negligible under our experimental conditions. The pump arm is modulated by an optical chopper at 12.5 kHz, and the pump-probe delay τ is set by a motorized translation stage in the pump arm, with $\tau > 0$ corresponding to the pump arriving first. Both beams are focused onto the sample by an off-axis parabolic mirror in a near-collinear geometry, with a mid-infrared beam waist radius at the sample of approximately $(46 \pm 4)$ µm determined by knife-edge measurements (see the Supplementary Materials for details). The sample is a [100]-oriented bulk germanium crystal of thickness $L = 0.5$ mm with optically polished surfaces. The linear refractive index of Ge at 5 µm is $n_0 \approx 4.0$ [48, 49], and the material is transparent at this wavelength.

The transmitted probe waveform is recorded by EOS [29–31]. For this, a small fraction of the 1030 nm fundamental is focused into a 6-mm-thick YAG plate to generate a broadband near-infrared supercontinuum, which is compressed to approximately 14 fs after dispersion compensation. The ultrashort gate is recombined collinearly with the probe and focused into a 200-µm-thick GaSe crystal. There the gate and the mid-infrared field undergo sum- and difference-frequency generation, and the newly generated components interfere linearly with the unconverted gate [50]. The result is a change in the ellipticity of the gate pulse that is proportional to the mid-infrared field. This polarization change is analyzed with a quarter-wave plate, a Wollaston prism, and a pair of balanced photodiodes. The full probe waveform $E(t)$ is mapped by sweeping the gate-probe delay $t$ with a piezo stage in a continuous sawtooth. Rapid-scan combined with position-triggered averaging suppresses slow temporal drifts between the gate and the mid-infrared pulse (see the Supplementary Materials).

Pump and probe are modulated at different subharmonics of the laser repetition rate to isolate the pump-induced change with high sensitivity. At each pump-probe delay $\tau$, the reference field $E_{ref}(t,\tau)$ and the pump-induced field change $\Delta E(t,\tau)$ are recovered simultaneously by demodulating the balanced-detector output at 25 kHz and 12.5 kHz, with a dual-channel lock-in amplifier. Demodulating the balanced output at both frequencies at once gives the unperturbed reference field $E_{ref}(t) \propto S_{25kHz} - S_{12.5kHz}$, together with the pump-induced change $\Delta E(t) \propto 2S_{12.5kHz}$; the relations and the factor of two are derived in the Supplementary Materials.

Recovering $E_{ref}$ and $\Delta E$ from the same measurement makes power fluctuations and slow path-length drifts common to both, so they cancel in $\Delta E/E_{ref}$, and the scheme reaches a relative sensitivity of $\approx 10^{-3}$ in the differential field. Note that the mid-infrared pulses are temporally chirped, and an uneven spectral response of the electro-optic detection, set by the crystal phase matching and the gate pulse, could imprint time-varying phase errors on the reconstructed waveform [51–53]; since $E_{ref}$ and $\Delta E$ traverse the same path, these errors are shared and drop out of $\Delta E/E_{ref}$, so the phase of the differential signal is preserved [39].

## RESULTS AND DISCUSSION

In a (degenerate) pump-probe geometry the intense pump modifies the refractive index experienced by the probe by

$$\Delta n = 2n_2 I_{pump} \tag{1}$$

where $n_2$ is the nonlinear refractive index, $I_{pump}$ is the pump intensity, and the factor of two is that of cross-phase modulation between co-polarized fields [54]. Over the sample thickness $L$ the probe accumulates a phase modulation

$$\Delta\varphi = \left(\frac{\omega_0}{c}\right)\Delta nL \tag{2}$$

where $\omega_0$ is the probe carrier frequency and $c$ is the speed of light. Note that both quantities follow the pump intensity envelope. Writing t for the EOS time, the field change recorded at each pump-probe delay τ is

$$\Delta E(t,\tau) = E_{ref}(t)\big[exp\big(i\Delta\varphi(t-\tau)\big) - 1\big]. \tag{3}$$

Figure 2(A) shows the measured field change $\Delta E(t,\tau)$ at each pump–probe delay, normalized to the peak reference field $E_{ref,max}$, at a peak pump intensity of $2.39 \pm 0.46$ GW cm$^{-2}$. The signal forms a ridge-like feature around $\tau = 0$, where pump and probe overlap in time, and falls below $1\times10^{-3}$ for delays beyond $\pm 250$ fs. The absence of any signal outside the overlap shows directly that no long-lived response, such as free-carrier absorption from photoinjected carriers, is generated at this intensity: the probe returns to its unperturbed form once the pulses temporally separate. The signal is tilted in the $(t,\tau)$ plane because the phase imprinted on the probe at sampling time $t$ follows the pump intensity at that instant, so the signal moves along $t$ as $\tau$ is scanned. The ridge, $t_{\mathrm{pk}}(\tau)$, is defined by the peak of the $\Delta E$ envelope at each pump–probe delay. A linear fit to the ridge points selected where the envelope amplitude exceeds 10% of its maximum within the delay range $|\tau| < 250$ fs, gives a slope of −0.66, close to the value −2/3 expected when the pump and probe are identical (see the Supplementary Materials for details). The amplitude threshold removes delays at which the pump and probe overlap only weakly, where the position of the envelope maximum becomes difficult to determine reliably because of the low signal amplitude, and the potentially non-Gaussian, chirped wings of the measured pulse; the delay range restricts the fit to the region of temporal overlap, within which the ridge is approximately linear.

A single slice at $\tau = 0$ reveals the character of the phase shift induced by the nonlinear Kerr effect. Figure 2(B) overlays $E_{ref}(t)$ (red) with the field change $\Delta E(t,\tau = 0)$ (blue), and the magnified inset compares their oscillations near the pulse center. The two waveforms oscillate at the same carrier frequency, while the pump-induced change is confined to the temporal overlap of the pump and probe pulses. Crucially, the two oscillations are not in phase: their zero crossings near $t = 0$ are displaced by 4.8 fs, a fraction 0.29 of the carrier period $T_C = 16.7$ fs at the 5 μm probe center wavelength. This near-quadrature relation between $\Delta E$ and $E_{ref}$ is the time-domain signature of a refractive response. In the small-phase limit, a refractive response gives $\Delta E \approx i\Delta\varphi E_{ref}$, so the field perturbation is 90° out of phase with the unperturbed probe field. In contrast, an absorptive response gives $\Delta E \propto -E_{ref}$, and therefore appears in antiphase.

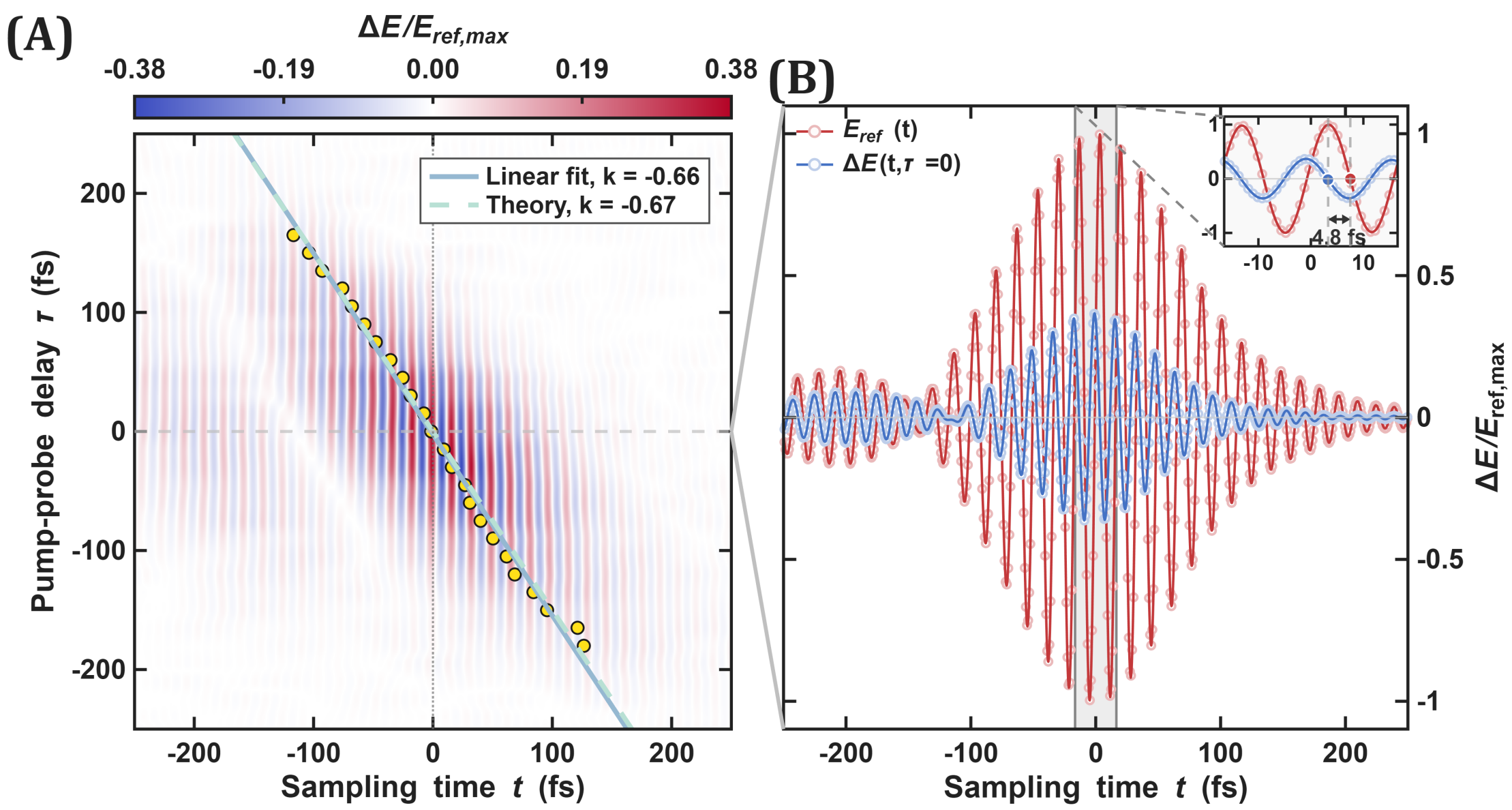

**Fig. 2.** Field-resolved cross-phase modulation in germanium in the time domain. (A) Pump-induced field change $\Delta E(t,\tau)$, normalized to the peak reference field $E_{\text{ref,max}}$, as a function of EOS time $t$ and pump-probe delay $\tau$, at a peak pump intensity of 2.39 GW $\text{cm}^{-2}$. The yellow markers are the maximum of the envelope of $\Delta E$, taken from the Hilbert transform along t. The signal forms a ridge-like feature confined to the pump-probe temporal overlap near $\tau = 0$ and falls below $1 \times 10^{-3}$ for $|\tau| > 250$ fs, showing that no long-lived response is generated at this intensity. The ridge is tilted in the $(t, \tau)$ plane because the phase imprinted at sampling time $t$ follows the instantaneous pump intensity; a linear fit to the ridge gives a slope of $-0.66$, close to the value $-2/3$ expected for equal gaussian pump and probe pulses. (B) Single delay slice at $\tau = 0$: reference field $E_{\text{ref}}(t)$ (red) and field change $\Delta E(t, \tau = 0)$ (blue). Inset: magnified view near $t = 0$; the marked interval is the 4.8 fs offset between the zero crossings.

Experimentally, the displacement lies slightly beyond 90°, as expected for a pure phase shift: the field change $\exp(i\Delta\varphi) - 1$ carries a phase $90° + \Delta\varphi/2$, thus a positive refractive $\Delta\varphi$ shifts the zero crossings just past quadrature. Since the displacement is dominated by the 90°quadrature term, it is only weakly sensitive to the magnitude of $\Delta\varphi$ and serves here as a qualitative signature of a refractive response; the quantitative value of $\Delta\varphi$ is obtained from the more detailed analysis below. The nonlinear response is encoded in the phase of the optical field, which is directly accessible from the recorded waveforms. For each waveform, we construct the analytic signal using a Hilbert transform, $A(t) = E(t) + i\mathcal{H}\{E(t)\}$, whose modulus gives the field envelope and whose argument gives the optical phase. At each pump-probe delay, the pump-induced phase change is evaluated at a sampling time t, according to $\Delta\varphi(t,\tau) = arg\left[1 + {A_{\Delta E}(t,\tau)}\big/{A_{E_{ref}}(t,\tau)}\right]$, where $A_{\Delta E}(t,\tau)$and $A_{E_{\text{ref}}}(t)$denote the analytic signals of the pump-induced field change and the unperturbed probe field, respectively. The phase response is evaluated within a $\pm 5$ fs window around probe-envelope maximum ($t = 0$) for all delays. This choice maximizes the pump-induced field change and thus provides a high signal-to-noise readout. Figure 3(A) shows the resulting phase response. It is centered near zero delay and rapidly returns to the baseline as the pump and probe separate. A delayed contribution with a finite lifetime, such as that associated with photogenerated carriers or lattice heating, would extend the response beyond the temporal overlap and produce a tail at positive delays. No such tail is resolved within the experimental noise, indicating that no significant delayed nonlinear contribution is observed under the present excitation conditions. A Gaussian fit gives a peak phase shift of $\Delta\varphi_0 = 0.27$ rad and a FWHM of 130 fs. Assuming identical pump and probe pulses, as supported by the ridge slope of −0.66 ≈ −2/3 (Fig. 2(A)), the measured 180 fs FWHM of the probe electric-field envelope corresponds to an intensity FWHM of 127 fs, consistent with the measured temporal width of the phase response. This behavior is consistent with a response dominated by the fast bound-electron contribution to the electronic $\chi^{(3)}$ nonlinearity [1].

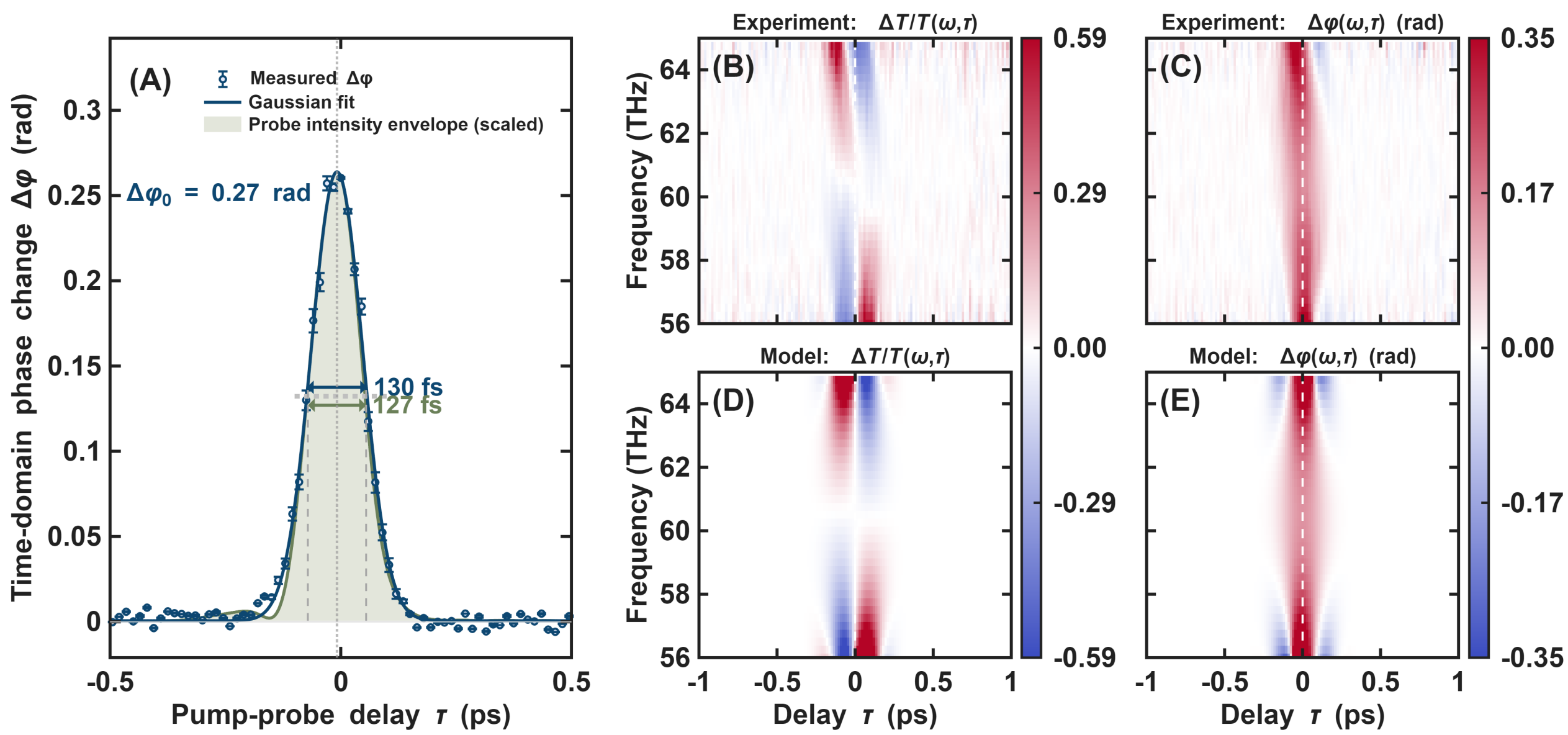

**Fig. 3.** Pump-induced phase response and its consistency with a phase-only model. (A) Pump-induced phase change evaluated at the peak of the probe-field envelope, $\Delta\varphi(\tau)$, as a function of pump-probe delay. The measured phase response is fitted with a Gaussian (blue), while the scaled probe intensity envelope is shown for comparison (green). The corresponding FWHMs are 130 fs for the phase response and 127 fs for the probe intensity envelope, indicating close temporal agreement. Error bars represent the standard deviation of the extracted phase obtained from sampling times within $| t | \leq$ 5fs. (B,C) Experimental fractional transmission change and phase change, respectively, across the probe band of 56–65 THz. (D,E) Corresponding quantities calculated from the phase-only forward model (see text). The model reproduces the main spectral and temporal features of the experimental transmission and phase responses.

To examine the spectral signature of the Kerr response, we Fourier transform the reference field $E_{ref}(t)$ and the pump-induced change $\Delta E(t,\tau)$ at each pump-probe delay, obtaining their frequency-domain signals, $\Delta\tilde{E}(\omega,\tau)$ and $\tilde{E}_{ref}(\omega)$. Defining $\tilde{r}(\omega,\tau) = \Delta\tilde{E}(\omega,\tau)/\tilde{E}_{ref}(\omega)$, where the tilde denotes a complex quantity in the frequency domain, the corresponding fractional transmission change and phase change are obtained as $\Delta T/T(\omega,\tau) = |1+\tilde{r}(\omega,\tau)|^2 - 1$, and $\Delta\varphi(\omega,\tau) = arg[1+\tilde{r}(\omega,\tau)]$. Figure 3(B) and (C) map these quantities across the probe band 56–65 THz over the full delay range. The phase change (Fig. 3(C)) remains predominantly positive across the probe band and is concentrated near zero delay, consistent with the time-domain response in Fig. 3(A). The fractional transmission change (Fig. 3(B)) exhibits both positive and negative spectral features concentrated around temporal overlap. This follows from the time dependence of the phase: a phase that varies across the probe pulse shifts its instantaneous frequency, redistributing spectral weight within the probe band. The direction of the shift depends on whether the probe overlaps the rising or falling edge of the pump intensity, so the sign of the spectral features reverses across zero delay.

To test whether this complex response originates from a pure phase modulation, we construct a forward calculation based on Eqs. (1)-(3) using Gaussian probe and pump pulses (127 fs intensity FWHM) and an instantaneous phase proportional to the pump intensity, with its peak fixed to the measured $\Delta\varphi_0 = 0.27$ rad. The pump-induced field change (Eq. (3)) is obtained by applying a time-dependent phase modulation whose temporal dependence and peak amplitude are determined from the experimentally measured $\Delta\phi_0$ in Fig. 3(A).

The corresponding transmission change $\Delta T/T(\omega,\tau)$ and phase change $\Delta\varphi(\omega,\tau)$ are then calculated from the simulated fields using the same procedure as for the experimental data, as shown in Figure 3(D) and (E). The model reproduces the principal features of both observables: the sign-changing spectral structure of $\Delta T/T(\omega,\tau)$ and the positive, single-signed phase response $\Delta\varphi(\omega,\tau)$ concentrated near zero delay. This agreement demonstrates that the phase response directly measured in the time domain captures the essential information required to reproduce the observed complex frequency-resolved response. Moreover, a frequency-dependent comparison is shown in the Supplementary Materials, where spectral cross-sections at representative delays reproduce the sign-changing transmission change and capture the main spectral trend of the phase response.

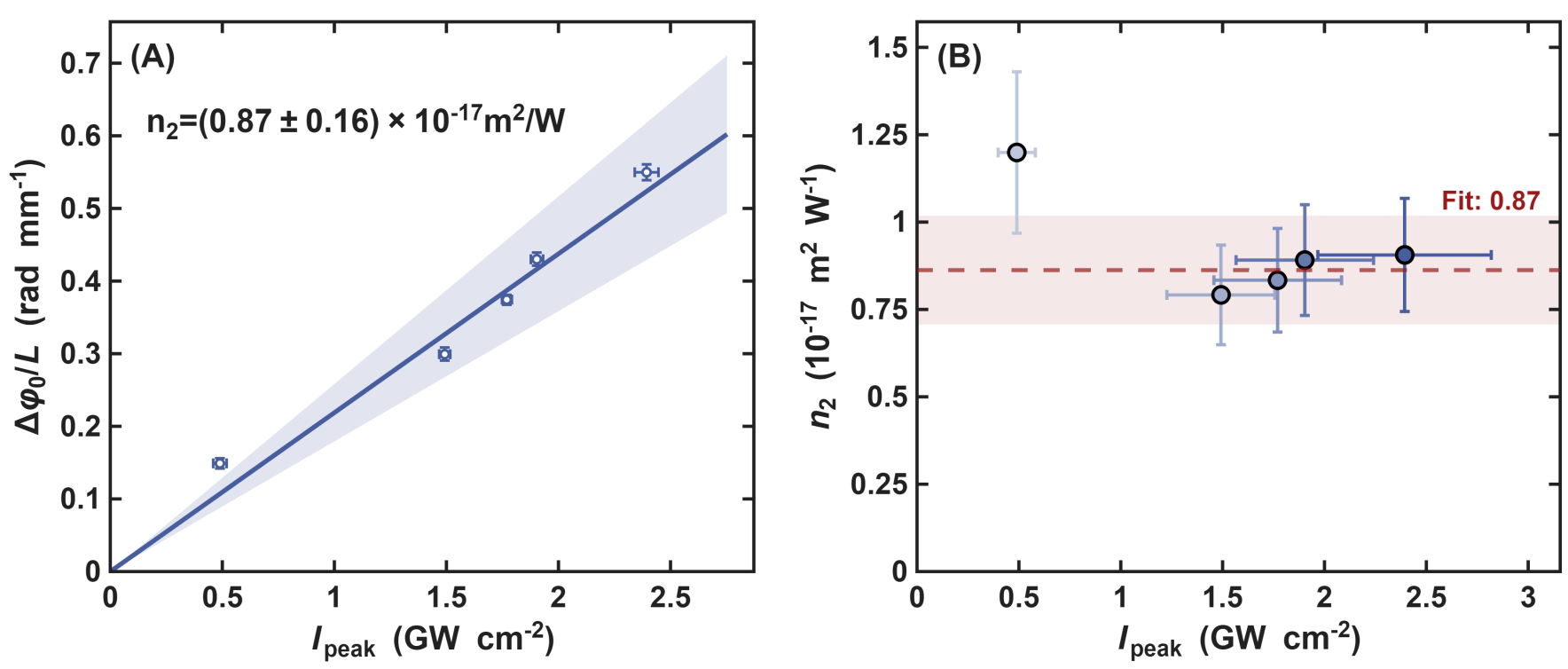

**Fig. 4.** Linear intensity scaling of the Kerr-induced phase shift and the nonlinear refractive index. (A) Peak phase shift per unit sample length $\Delta\varphi_0/L$ versus in-sample peak intensity $I_{\mathrm{peak}}$, for five pump intensities each averaged over three independent scans. The data follow the linear fit with no offset and no curvature over the full range, the scaling expected for an instantaneous Kerr nonlinearity. The shaded region represents the uncertainty band of the linear fit, accounting for the statistical and systematic uncertainties in $n_2$. (B) Nonlinear refractive index $n_2$ extracted independently at each pump intensity. The dashed line indicates the globally fitted value of $n_2$, and the shaded region shows the same uncertainty band obtained from the linear fit in (A).

The refractive nature of the response is further supported by its linear scaling with pump intensity. For each of the five pump intensities we extract the peak phase $\Delta\varphi_0$ using the procedure described for Fig. 3(A). Figure 4(A) shows the dependence of $\Delta\varphi_0$ normalized to the sample length on the in-sample peak intensity $I_{peak}$. The data exhibit an approximately linear dependence and are well described by a linear fit constrained to zero intercept. The slope of the fit gives the nonlinear refractive index directly through Eqs. (1) and (2), and yields $n_2 = (0.87 \pm 0.16) \times 10^{-17}\mathrm{m}^2/\mathrm{W}$, where the uncertainty includes both statistical and systematic contributions. This value is of the same order as the mid-infrared $n_2$ of germanium previously obtained by Z-scan ($\sim 5 \times 10^{-18}$ m$^2$/W near 5 μm) [45]. The vertical error bars in Fig. 4(A) represent the standard error of the mean from three independent measurements at each pump intensity, while the horizontal error bars denote the short-term fluctuations of the measured pump power. The systematic uncertainties of the pump-beam radius and the power-meter calibration rescale all intensities by a common factor and therefore do not affect the point-to-point scatter. They are instead included in the shaded region, which represents the range of fitted lines consistent with the total uncertainty. The values of $n_2$ extracted independently at each intensity are shown in Fig. 4(B), with error bars representing the combined statistical and systematic uncertainties.

The linear scaling in Fig. 4(A) is expected for an instantaneous Kerr response, for which the pump-induced cross-phase modulation phase shift is proportional to the pump intensity. Refraction by carriers generated through three-photon absorption, the lowest-order absorption process at 5 μm, would scale with the cube of the intensity and cause a deviation from linear scaling, which is not observed within the uncertainties. Together with the absence of a delayed tail in Fig. 3(A), this indicates that the response is governed by the instantaneous Kerr nonlinearity.

## CONCLUSION AND OUTLOOK

We have characterized the Kerr nonlinear response of germanium at 5 μm by field-resolved degenerate pump probe spectroscopy, recording the full transmitted electric field waveform and thus the transmission change $\Delta T/T$ rather than an intensity only. Under our experimental conditions, the measured phase response follows the pump intensity envelope and vanishes when the pump and probe no longer overlap, consistent with an instantaneous Kerr response. The phase change scales linearly with pump intensity, and from this scaling we obtain a nonlinear refractive index $n_2 = (0.87 \pm 0.16) \times 10^{-17} \mathrm{m}^2/\mathrm{W}$. By directly resolving the probe electric field in the time domain, our approach provides direct access to the pump-induced phase response and its evolution with pump-probe delay, rather than relying on intensity changes as the primary observable. Because pump and probe share the same spectrum, the coefficient obtained is the degenerate one, without conversion from a non-degenerate value, and the two pulses propagate at the same group velocity so that no walk-off arises between them. The approach is not specific to germanium or to 5 μm and could be extended to other semiconductors, such as silicon, provided that the material is transparent to the probe at the chosen wavelength. Tuning the degenerate wavelength would extend the same measurement across the mid-infrared transparency range, where data on the dispersion of the Kerr index remain scarce.

In this work the pump intensity was held below the threshold for carrier injection, isolating the purely refractive Kerr response. Raising the intensity across this threshold would inject real carriers, through multiphoton absorption and, at sufficiently strong fields, through tunnelling and the dynamical Franz–Keldysh effect. The response would evolve from the instantaneous Kerr behavior reported here toward a free-carrier contribution with a finite buildup and lifetime. This crossover between the bound-electron and free-carrier nonlinearities that ultimately bound the usable refractive response of a driven semiconductor. This strong-field regime is precisely where field-resolved detection has proven powerful on the carrier side: the complex dielectric response of photoinjected carriers, and most recently the impact-ionization and trapping dynamics of hot charges in silicon and silica, have been followed directly in the recorded waveform [40]. Since the present scheme records the same complex observable below the injection threshold, it provides the refractive counterpart to those measurements, and together the two would map how a semiconductor crosses from a purely refractive to a dissipative nonlinearity within a single framework.

## ACKNOWLEDGMENTS

**Author contributions:** H. L. and N. M. conceived the idea. H. L, H. K. and H. B. designed and performed the experiments. H. L. performed the theoretical calculations with input from H. K., N. M., A. D. A. and M. L. N.M. supervised the research. All authors contributed to the data analysis, discussions and the writing of the manuscript.

**Funding:** This work was supported by the Swedish Research Council (Grants No. 2021-05784 and No. 2025-04734), the Knut and Alice Wallenberg Foundation (Grant No. 2023.0089), the European Innovation Council Pathfinder Open project “iSenseDNA” (Grant No. 101046920), the European Research Council (ERC Starting Grant No. 101116253 “MagneticTWIST”), Kempestiftelserna (Grant No. JCK-3122) and HORIZON EUROPE Marie Sklodowska-Curie Actions (Grant No. 101081455).

**Competing interests:** The authors declare that they have no competing interests.

## DATA AVAILABILITY

All data supporting the findings of this article is available upon reasonable request.

## Supplementary Material

Beam quality ($M^2$) characterization

The spatial quality of the 5 μm probe beam at the sample plane was characterized by the knife-edge technique combined with a caustic ($M^2$) measurement following ISO 11146. At several longitudinal positions along the focusing axis, a razor edge mounted on a manual translation stage was scanned transversely across the beam while the transmitted power was recorded. For a beam with a Gaussian transverse intensity profile, the transmitted power follows an error function

$$P(x) = P_0 + \frac{P_{\max}}{2}\left[1 - \operatorname{erf}\left(\sqrt{2}\,\frac{x - x_0}{w}\right)\right],$$

where $P_0$ is the background power (detector offset and stray light) measured when the knife edge fully blocks the beam, $P_{\max}$ is the additional power recovered when the edge is fully retracted, $x_0$ is the beam centroid and $w$ is the $1/e^2$ intensity radius. The four parameters $\{P_0, P_{\max}, x_0, w\}$ were extracted at each longitudinal position by nonlinear least-squares fitting. A representative scan and fit along the measured transverse direction are shown in Fig. S1(A). The measured profile is reasonably well described by a Gaussian function, with the fit residuals ($< 0.1\%$ of $P_{\max}$, RMS $\approx$ 52μW). The resulting beam radii were fit to the standard hyperbolic caustic

$$w^2(z) = w_0^2\left[1 + \left(\frac{z - z_0}{z_R}\right)^2\right],\ z_R = \frac{\pi w_0^2}{M^2\lambda}$$

with $\lambda = 5$μm. The global caustic fit (Fig. S1(B)) yields a beam-waist radius of $w_0 = 50.1 \pm 2.5\ \mu$m, a Rayleigh range of $z_R = 1.20 \pm 0.08$ mm, a far-field half-divergence of $41.6 \pm 1.6$ mrad, and a beam-propagation factor $M^2 = 1.31 \pm 0.07$.

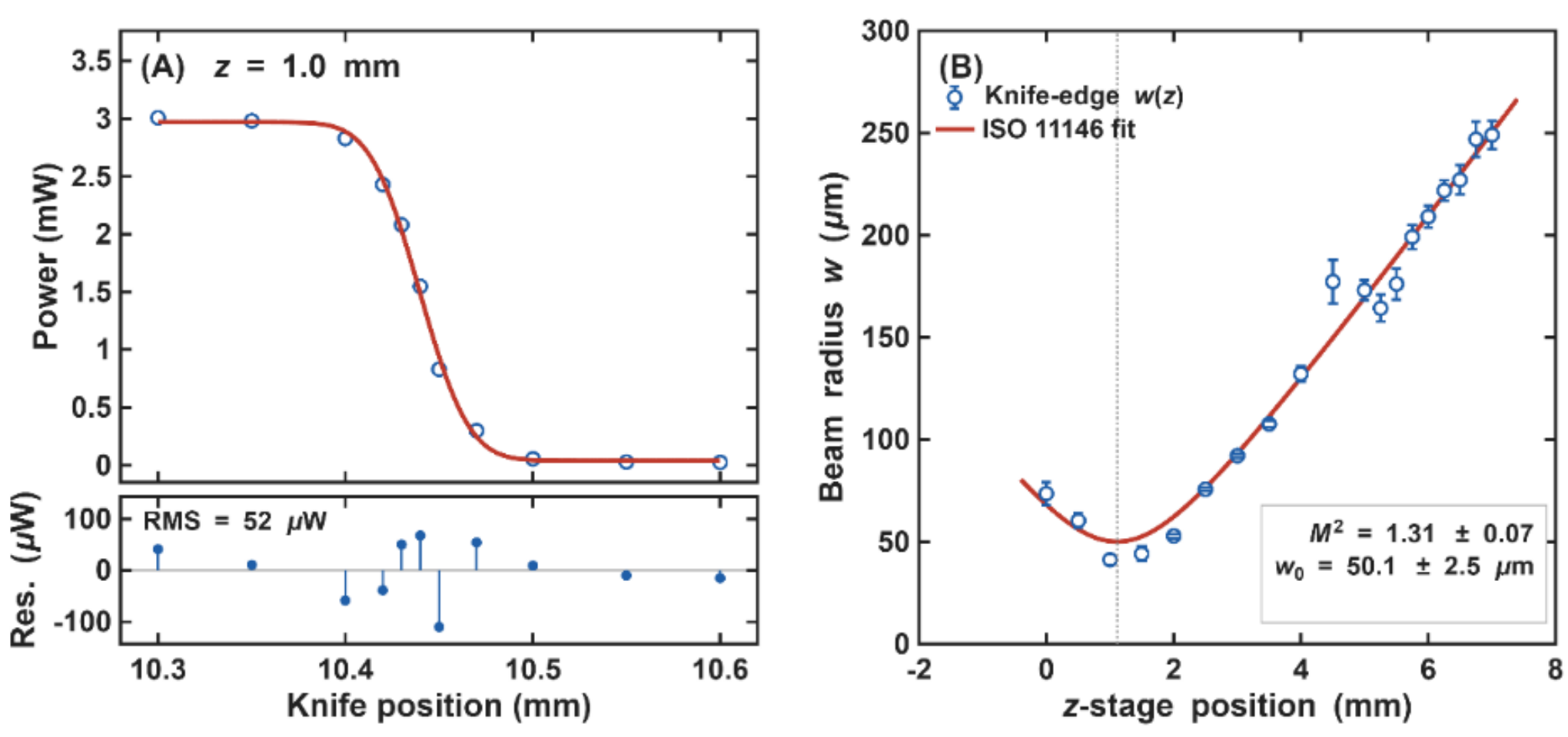

**Fig. S1.** Spatial characterization of the 5 μm probe beam. (A) Representative knife-edge power scan (open circles) at $z = 1$ mm with the error-function fit (solid red line); the lower strip shows the fit residuals. The extracted $1/e^2$ radius at this position is $w =$ 42 μm. (B) Beam radius $w$ as a function of stage position $z$ (open circles) with the ISO 11146 hyperbolic-caustic fit (solid red line); the dotted line marks the waist position $z_0$. Vertical error bars are the $1\sigma$ statistical standard errors of $w$ from the covariance matrix of the individual error-function fits and reflects fit statistics only (systematic contributions from stage positioning and detector response are not included).

The near-unity $M^2$ confirms a close-to-diffraction-limited focus and justifies the Gaussian spatial profile assumed in the peak-intensity calibration. The fit yields a beam-waist radius of approximately 50 $\mu$m, whereas the knife-edge measurements near the focus indicate a local beam radius of approximately 42 $\mu$m. This difference suggests that the uncertainty in the beam radius relevant to the peak-intensity calibration is not fully represented by the statistical uncertainty of the global fit alone. We therefore use an effective beam radius of $w_0 = 46 \pm 4\ \mu$m for the peak-intensity calculation. This value lies between the global-fit result and the near-focus measurements, and the assigned uncertainty is intended to encompass the difference between the two estimates. The resulting uncertainty in the beam radius is propagated into the peak pump intensity and the extracted nonlinear refractive index.

Dual-modulation detection and rapid-scan field sampling

Field-resolved cross-phase-modulation spectroscopy requires accurate measurement of both the transmitted probe field $E_{ref}$ and the pump-induced field change $\Delta E$, while minimizing the effects of laser fluctuations and slow experimental drifts. We therefore employ two complementary techniques, summarized in Fig. S2: a dual-modulation scheme that encodes the probe field and its pump-induced change on separate lock-in channels (Fig. S2(A)), and a rapid-scan sampling of the electro-optic gate delay that renders the measurement immune to slow drift (Fig. S2(B)). Together, they improve the stability and reproducibility of the field-resolved measurement. The two are described in turn below.

Dual-modulation demodulation. The mid-infrared output is generated by transmitting every second pulse of the $f_{\text{rep}} = 50$ kHz oscillator through a Pockels cell, giving a MIR repetition rate of $f_{\text{rep}}/2 = 25$ kHz (see Fig. 1(B) in main text). The pump arm is then gated by a mechanical chopper synchronized to $f_{\text{rep}}/4 = 12.5$ kHz, so that every second MIR pump pulse is blocked (Fig. S2(A)). The two modulation frequencies are therefore harmonically related to the laser repetition rate, with a fixed phase relation in the pulse sequence. Indexing the pulses at the original 50 kHz repetition rate, one modulation period at 12.5 kHz contains four pulse slots. The detected probe field is then

$$\left[E_{ref} + \Delta E, 0, E_{ref}, 0\right],$$

where $E_{ref}$ is the transmitted probe field without the pump and $\Delta E$ is the pump-induced field change. The corresponding four-point discrete Fourier coefficients from the pulse sequence above are defined as

$$a_m = \frac{1}{4}\sum_{n=0}^{3} S_n e^{-i2\pi mn/4}$$

Here, $S_n$ denotes the detected probe-field signal in the $n$-th pulse slot. For the two lock-in frequencies used in the experiment, substitution gives

$$a_1 = \frac{\Delta E}{4},\ a_2 = \frac{2E_{ref} + \Delta E}{4}$$

corresponding to $f_{\text{rep}}/4$, and $f_{\text{rep}}/2$, respectively. Thus, using the same amplitude convention for the two lock-in channels

$$S_{f_{\text{rep}}/4} \propto \frac{\Delta E}{2},\ S_{f_{\text{rep}}/2} \propto E_{ref} + \frac{\Delta E}{2}$$

The reference field and pump-induced field change are consequently reconstructed as

$$E_{ref} \propto S_{frep/2} - S_{frep/4}, \Delta E \propto\ 2S_{frep/4},$$

and hence

$$\frac{\Delta E}{E_{ref}} = \frac{2S_{frep/4}}{S_{frep/2} - S_{frep/4}}$$

Rapid-scan acquisition. Each field transient is recorded in a single 0.5 s half-sweep of the piezo-driven gate delay (red in Fig. S2(B)), rather than by dwelling at a fixed delay position. A position trigger is recorded for each sweep to identify the scan boundaries and segment the continuous acquisition into individual transients. The drift that matters here is the variation of the relative arrival time between the mid-infrared field and the gate pulse, which shifts successive waveforms along the sampling time axis. This timing offset is determined for each transient by cross-correlation with a reference waveform, and the individual transients are shifted accordingly before averaging. The position trigger therefore defines the individual scan cycles, while the cross-correlation provides the timing-drift correction. Because a sweep is completed in a time short compared with the drift, all delay points within one transient are acquired under the same conditions, and repeated, temporally aligned sweeps can be averaged to improve the signal-to-noise ratio. Together with the dual-modulation detection described above, this acquisition scheme reduces the influence of laser fluctuations and slow experimental drifts.

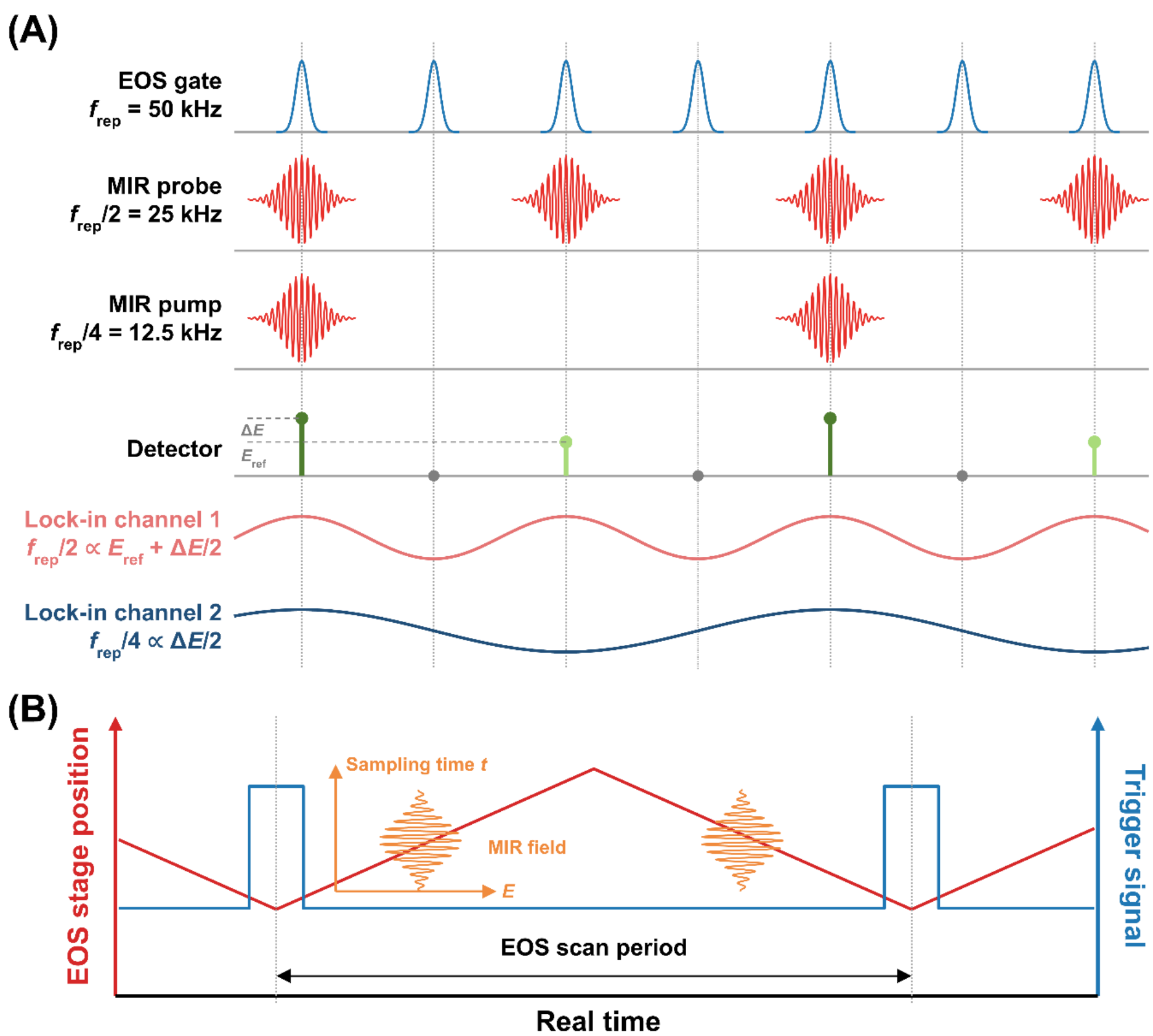

**Fig. S2.** Dual-modulation detection and rapid-scan field sampling. (A) Pulse-picking and demodulation scheme. The mid-infrared output is modulated at the first subharmonic of the repetition rate ($f_{rep}/2$ = 25 kHz) by a Pockels cell (see Fig. 1(B) in main text) and then split into pump and probe arms; the pump arm is modulated a second time at the second subharmonic ($f_{rep}/4$ = 12.5 kHz) by a mechanical chopper. Every gate pulse is measured by the detector, so over one 12.5 kHz period the four readings are $[E_{ref} + \Delta E, 0, E_{ref}, 0]$ (stem plot). Lock-in demodulation at 25 kHz yields $E_{ref} + \Delta E/2$ and at 12.5 kHz yields $\Delta E/2$, from which $E_{ref}$ and $\Delta E$ are obtained. (B) Rapid-scan acquisition. The electro-optic gate delay is swept along a periodic trajectory (red) so that a complete mid-infrared field transient (orange) is recorded against sampling time t on every 0.5 s half-sweep. A synchronized position trigger (blue) marks the sweep boundaries and segments the continuous record into individual transients. Each transient is then aligned to a common reference by cross-correlation before averaging. The horizontal axis is laboratory (real) time; the sampling time axis t runs along each sweep.

Ridge slope in the degenerate limit

The ridge slope provides a consistency check of the temporal structure of the measured cross-phase modulation response and its expected behavior for degenerate pump and probe pulses. In the two-dimensional map shown in Fig. 2(A), the amplitude envelope of the cross-phase-modulation induced field change forms a ridge. For an instantaneous XPM response, the ridge position is determined by the temporal overlap of the probe field envelope and the pump intensity envelope. For an idealized case of identical Gaussian pump and probe pulses, the expected slope is $-2/3$, as derived below. Because the pump and probe originate from the same mid-infrared source and traverse the same dispersion, both experience an identical group-delay dispersion $D_2$ and hence the same chirp-induced stretching; this common factor is what ultimately drops out of the slope.

*Step 1 — chirp-induced broadening.*

A transform-limited Gaussian pulse of duration $\tau_{TL}$ acquires, under a group-delay dispersion $D_2$, a stretched duration $\tau_{\tilde{E},c} = \tau_{TL} \cdot S$, with the common stretching factor

$$S = \sqrt{1 + (\frac{D_2}{\tau_{TL}{}^2})^2} \geq 1.$$

The chirp stretches the field and intensity envelopes by the same factor S. The relevant widths are that of the probe field envelope, $\tau_{\tilde{E},c}$, and that of the pump intensity envelope, $\tau_{I,c}$.

*Step 2 — ridge position.*

The ridge follows the delay $t$ at which the differential field envelope is maximal. Since $\Delta E(t,\tau)$ is proportional to the product of the (Gaussian) chirped probe field envelope at $t$ and the chirped pump intensity envelope at $t + \tau$,

$$|\Delta E(t,\tau)| \propto exp(-\frac{t^2}{\tau_{\tilde{E},c}{}^2}) \cdot exp(-\frac{(t+\tau)^2}{\tau_{I,c}{}^2}),$$

setting $\partial|\Delta E|/\partial t = 0$ gives

$$-\frac{2t}{\tau_{\tilde{E},c}{}^2} - \frac{2(t+\tau)}{\tau_{I,c}{}^2} = 0,$$

which solves for the ridge position

$$t_{ridge}(\tau) = -\tau \cdot \frac{\tau_{\tilde{E},c}{}^2}{\tau_{I,c}{}^2 + \tau_{\tilde{E},c}{}^2}$$

The ridge slope is therefore

$$k = \frac{dt_{ridge}}{d\tau} = -\frac{\tau_{\tilde{E},c}{}^2}{\tau_{I,c}{}^2 + \tau_{\tilde{E},c}{}^2}$$

*Step 3 — degenerate case.*
For degenerate pump and probe with identical dispersion, the probe field envelope and the pump intensity envelope stretch to

$$\tau_{\tilde{E},c} = \sqrt{2} \cdot \tau_{TL} \cdot S, \tau_{I,c} = \tau_{TL} \cdot S,$$

For Gaussian pulses, the intensity envelope is narrower than the field envelope by a factor of $\sqrt{2}$. Substituting, the transform-limited duration and the stretching factor S cancel completely between numerator and denominator:

$$k = -\frac{(\sqrt{2}\tau_{TL}S)^2}{(\tau_{TL}S)^2 + (\sqrt{2}\tau_{TL}S)^2} = -\frac{2\tau_{TL}{}^2S^2}{3\tau_{TL}{}^2S^2} = -\frac{2}{3}$$

The slope of −2/3 is thus independent of the applied dispersion $D_2$, because the common stretching factor cancels. For Gaussian pulses with identical pump and probe temporal durations, this model predicts a ridge slope of −2/3. The experimentally measured slope is close to this value, with the small deviation potentially arising from differences between the actual temporal profiles and the ideal Gaussian model. Since the slope depends only on the ratio of the pump and probe durations, the measured value of −0.66 is consistent with nearly identical pump and probe durations. This supports the assumption of identical pump and probe pulses used in the following analysis, as the pump waveform is not measured directly.

Spectral cross-sections of the complex transmission change
Figure S3 compares the measured transmission change $\Delta T/T(\omega,\tau)$ and phase change $\Delta\varphi(\omega,\tau)$ with the pure-phase forward model of the main text, plotted as spectral cross-sections at three representative pump-probe delays ($\tau = -0.1, 0, +0.1$ ps) of the full two-dimensional maps of Fig. 3. Points are experimental data; solid lines are the model.
Figure S3(A) shows the transmission change $\Delta T/T(\omega,\tau)$ across the probe band at three representative delays. At $\tau = \pm 0.1$ ps, the spectrum is antisymmetric about the band centre, with a zero crossing near 60 THz, and its sign reverses between the two delays. At $\tau = 0$ the transmission change is much weaker. The model reproduces the antisymmetric line shape at $\pm 0.1$ ps and the weak response at zero delay. As discussed for Fig. 3(B) in the main text, the time-dependent phase shifts the instantaneous frequency of the probe. At $\tau = \pm 0.1$ ps the probe overlaps mainly the rising or the falling edge of the pump, so its spectrum is shifted towards one side, and the direction reverses with the sign of the delay. At $\tau = 0$ the probe overlaps both edges symmetrically, and the opposite shifts largely cancel.

Figure S3(B) shows the corresponding phase change $\Delta\varphi(\omega, \tau)$, which remains positive across the spectrum and is largest near zero delay. The model captures this sign and its spectral trend. At zero delay the measured amplitude agrees with the model near the center of the band but falls below it towards the band edges, where the reference spectral amplitude in the denominator becomes small and the ratio is correspondingly more sensitive to noise. The comparison is therefore quantitative for the spectral structure of $\Delta T/T$ and qualitative for the absolute phase amplitude; in both cases the model reproduces the character of the response without any absorptive term.

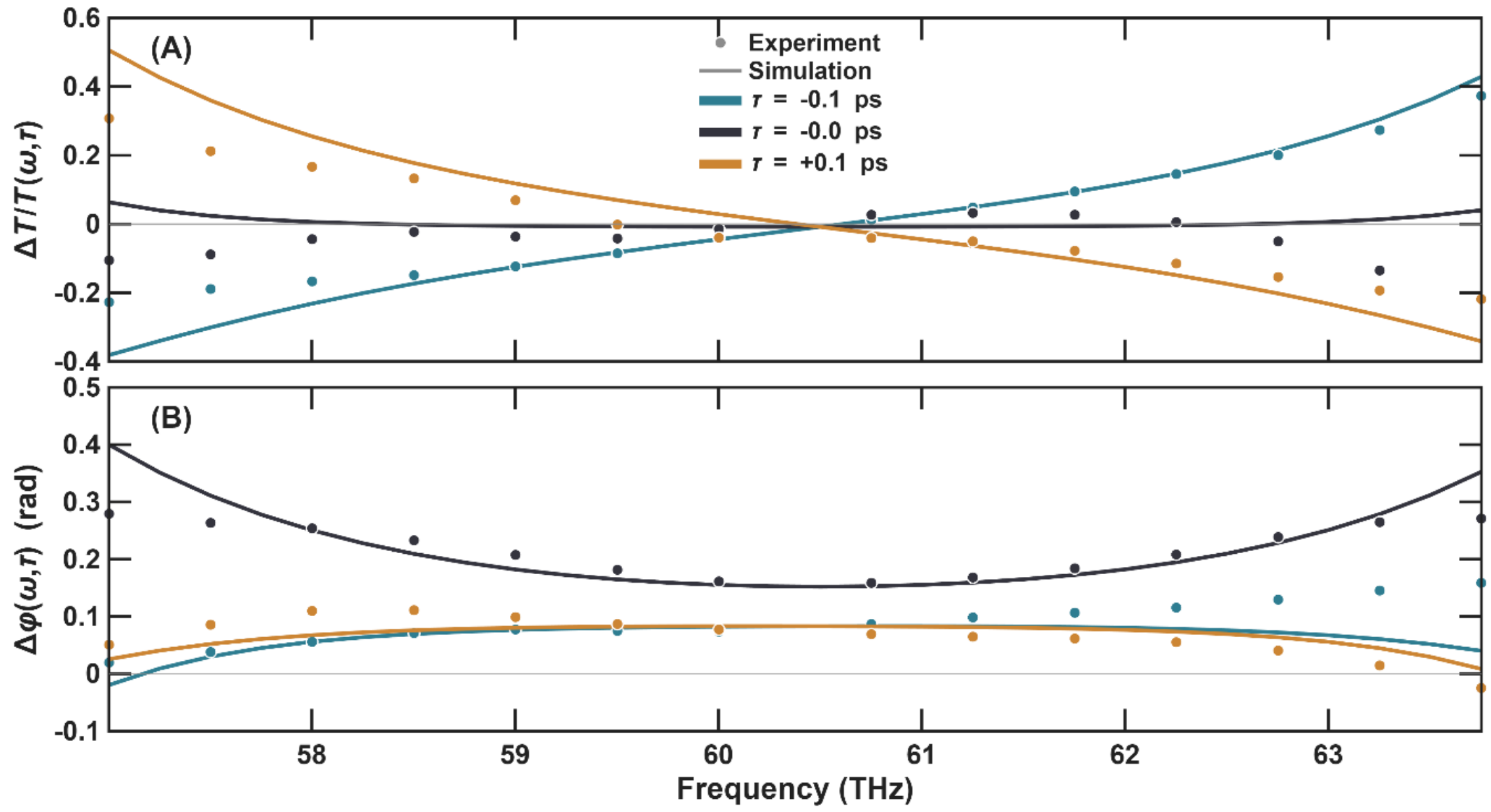

**Fig. S3.** Spectral cross-sections of the transmission change and phase changes shown in Fig. 3(B,C) and Fig. 3(D,E). Fractional transmission change (A) and phase change (B) as a function of frequency across the probe band, at pump–probe delays $\tau = -0.1$ ps (teal), 0 ps (black), and $+0.1$ ps (amber). Points are experiment; solid lines are the pure-phase forward model of Fig. 3. The model reproduces the sign-changing, delay-reversing spectral structure of the fractional transmission change (panel (A) for $\tau = \pm 0.1$ ps) and captures the single-signed phase response peaked near zero delay (panel (B)).